\documentclass[twocolumn,prl,showpacs,english,aps]{revtex4-1}
\usepackage{babel,rotating,dcolumn}
\usepackage{colordvi,graphicx,color,amsbsy,amsmath,bm,amssymb}
\usepackage{ulem} 
\usepackage{comment}
\usepackage{mathrsfs}

\newcommand*\revision[1]{\textcolor{black}{#1}}

\begin{document}
\title[]
{Heavy Particle Clustering in Inertial Subrange of High--Reynolds Number Turbulence }
\author{Keigo Matsuda$^{1}$} 
\email{k.matsuda@jamstec.go.jp}
\author{Katsunori Yoshimatsu$^{2}$} 
\author{Kai Schneider$^{3}$} 
\affiliation{$^{1}$ Research Institute for Value-Added-Information Generation (VAiG), Japan Agency for Marine-Earth Science and Technology (JAMSTEC), Yokohama 236-0001, Japan}
\affiliation{$^{2}$ Institute of Materials and Systems for Sustainability, Nagoya University, Nagoya, 464-8601, Japan}
\affiliation{$^{3}$ Institut de Math\'ematiques de Marseille (I2M), Aix-Marseille Universit\'e, CNRS, 13331 Marseille Cedex 3, France}
\date{\today}
\begin{abstract}
Direct numerical simulation of homogeneous isotropic turbulence shows pronounced clustering of inertial particles in the inertial subrange at high Reynolds number, in addition to the clustering typically observed in the near dissipation range.
The clustering in the inertial subrange is characterized by the bump in the particle number density spectra and is due to modulation of preferential concentration.  
The number density spectrum can be modeled by a rational function of the scale-dependent Stokes number.
\end{abstract}

\maketitle
%
Inertial heavy particles suspended in turbulence are frequently observed in geophysical and industrial flows, such as cloud droplets and volcanic ash in atmospheric turbulence, dust particles in protoplanetary disks, and fuel droplets in spray combustion. 
The particles show nonuniform distribution, i.e., clustering, in turbulence due to deviation of particle motion from the fluid motion \cite{Maxey,Squires&Eaton1990,Squires}.
Clustering of inertial heavy particles has been studied extensively, for reviews we refer to, e.g., Refs.~\cite{Shaw, Toschi, Monchaux2012, Brandt&Coletti2022, BGM2024}.
The clustering can be characterized statistically by two-point correlation of number densities such as the radial distribution function \revision{(RDF)} and the number density spectrum. 
The former is relevant to, e.g., droplet collision and coalescence statistics in raindrop formation process \cite{OV2014,Ireland2016}, and the latter gives estimates of clustering influence on cloud radar reflectivity factor \cite{Matsuda2014,MO2019}.
The mechanism of the clustering is explained by the preferential concentration in the pioneering work by Maxey \cite{Maxey} 
when the particle relaxation time $\tau_p$ is sufficiently smaller than the flow time scale: 
inertial particles are swept out from turbulent vortices due to the centrifugal effect and concentrate in low vorticity and high strain-rate regions. 
The inertial effect can be quantified by the Stokes number defined as $St \equiv \tau_p/\tau_\eta$ with the Kolmogorov time $\tau_\eta \equiv \sqrt{\nu/\epsilon}$. Here, $\nu$ is the kinematic viscosity, and $\epsilon$ is the mean energy dissipation rate per unit density.
The review \cite{Brandt&Coletti2022} well summarize clustering mechanisms including other ones proposed for larger $\tau_p$.

Inertial particles show multiscale clustering structures \cite{Boffetta2004,GV2006,YG2007,Monchaux2010,Bragg2015,Ariki2018,OG2021}, 
whose number density spectrum has a pronounced bump in the near dissipation range, i.e., the scales between the inertial and dissipation ranges \cite{Matsuda2014}.
The closure analysis in Ref.~\cite{Ariki2018} predicted a universal scaling for the number density fluctuation in the inertial subrange. To observe such inertial subrange clustering, one needs sufficiently high--Reynolds number turbulence so that the spectrum of fluid velocity and pressure obeys the scaling based on Kolmogorov's idea \cite{K41} (K41), according to the theories in Refs.~\cite{Bragg2015,Ariki2018}.
The recent development of supercomputers enables us to explore the influence of the Reynolds number on the clustering by using direct numerical simulation (DNS) of particle-laden turbulence.
In this letter, we examine the inertial subrange clustering and its dependence on the Stokes number by performing three-dimensional DNS at high Reynolds number.

We consider statistically homogeneous turbulent flow governed by the incompressible Navier--Stokes (N--S) equation,  
$\partial {\bm u} /\partial t + {\bm u}{\bm \cdot}\nabla{\bm u}  = -\nabla {\cal P} +\nu \nabla^2 {\bm u} + {\bm f}$,
where the velocity ${\bm u}({\bm x},t)$ satisfies $\nabla{\bm \cdot}{\bm u} = 0$. 
Pressure per density is denoted by ${\cal P}({\bm x},t)$, and ${\bm f}({\bm x},t)$ is an external solenoidal forcing. 
We use a cubic domain with length $2 \pi$ and periodic boundary conditions.
The particle size is assumed to be sufficiently smaller than the Kolmogorov length $\eta \equiv (\nu^3/\epsilon)^{1/4}$, 
and $\rho_p/\rho \gg 1$, where $\rho_p$ and $\rho$ are the particle and fluid densities, respectively.
Then, the Lagrangian motion of inertial heavy particles is governed by $d {\bm x}_p/d t = \revision{{\bm v}_p}$ and $d \revision{{\bm v}_p}/d t = -\{\revision{{\bm v}_p} - {\bm u}({\bm x}_p)\}/\tau_p$,
where ${\bm x}_p$ and \revision{${\bm v}_p$} are the position and velocity of a Lagrangian particle, respectively \cite{Squires}.
The Taylor-microscale Reynolds number is defined as $Re_\lambda \equiv u'\lambda/\nu$, where $u' \equiv \sqrt{\langle|{\bm u}|^2\rangle/3}$ is the turbulent velocity fluctuation, 
and $\lambda \equiv \sqrt{15 \nu u'^2/\epsilon}$ is the Taylor microscale.  
Here, $\langle\cdot\rangle$ denotes an ensemble average. 
%
A series of DNS computations of particle-laden turbulence was performed using the same numerical code as in Refs.~\cite{Onishi2011,Matsuda2014,MO2019,K3}.
External solenoidal forcing was applied at large scale, $|{\bm k}|<2.5$, where ${\bm k}$ is the wave number vector, to obtain statistically stationary turbulence, following a random forcing scheme in Ref.~\cite{YA2007} (see details in the Supplemental Material (SM) \cite{SM}).
Inertial particles were seeded uniformly and randomly in the turbulent flow at a statistically steady state, defined as $t=0$. 
Particle position data were then sampled at 10 time instants of $t=11 T_0$ to $20 T_0$ at interval of $T_0$, where $T_0$ is the dimensionless time unit, comparable to the large-eddy turnover time.
Spectra in this letter are averaged for the data at the sampling time instants.
The DNS parameters and the turbulent flow statistics are summarized in Table~\ref{tab:DNS}. 
The ensemble average for the statistics is computed as spatial and temporal average.
The latter is taken for the period of $10T_0 \le t \le 20T_0$.
The Stokes number $St$ is 0.05, 0.1, 0.2, 0.5, 1.0, 2.0 and 5.0. 

\begin{table}[t!]
\begin{center}
\begin{tabular}{c ccccccc}
\hline\hline
$N_g$ & $N_p$ & $Re$ & $Re_\lambda$ & $u'$ & $\epsilon$ & $k_{\mathrm{max}}\eta$ \\
\hline
512  & 1.5 $\times 10^7$ & 816   & 155 & 0.97 & 0.451 & 2.05 \\
1024 & 5.0 $\times 10^7$ & 2052  & 251 & 0.97 & 0.438 & 2.07 \\
2048 & 4.0 $\times 10^8$ & 5257  & 402 & 0.96 & 0.417 & 2.07 \\
4096 & 3.2 $\times 10^9$ & 13212 & 648 & 0.99 & 0.453 & 2.03 \\
\hline\hline
\end{tabular}
\end{center}
\caption{DNS parameters and statistics of obtained turbulence; 
the number of grid points $N_g$, the number of particles $N_p$, 
the Reynolds number of DNS, \revision{$Re = U_0 L_0/\nu$}, 
$Re_\lambda$, $u'$, $\epsilon$, and $k_{\mathrm{max}} \eta$, where $k_{\rm max} \equiv N_g/2$ is the maximum wave number\revision{, and $U_0$ and $L_0$ are unity}.
}
\label{tab:DNS}
\end{table}

\begin{figure}[tb!]
    \centering
    \includegraphics[width=8.6cm]{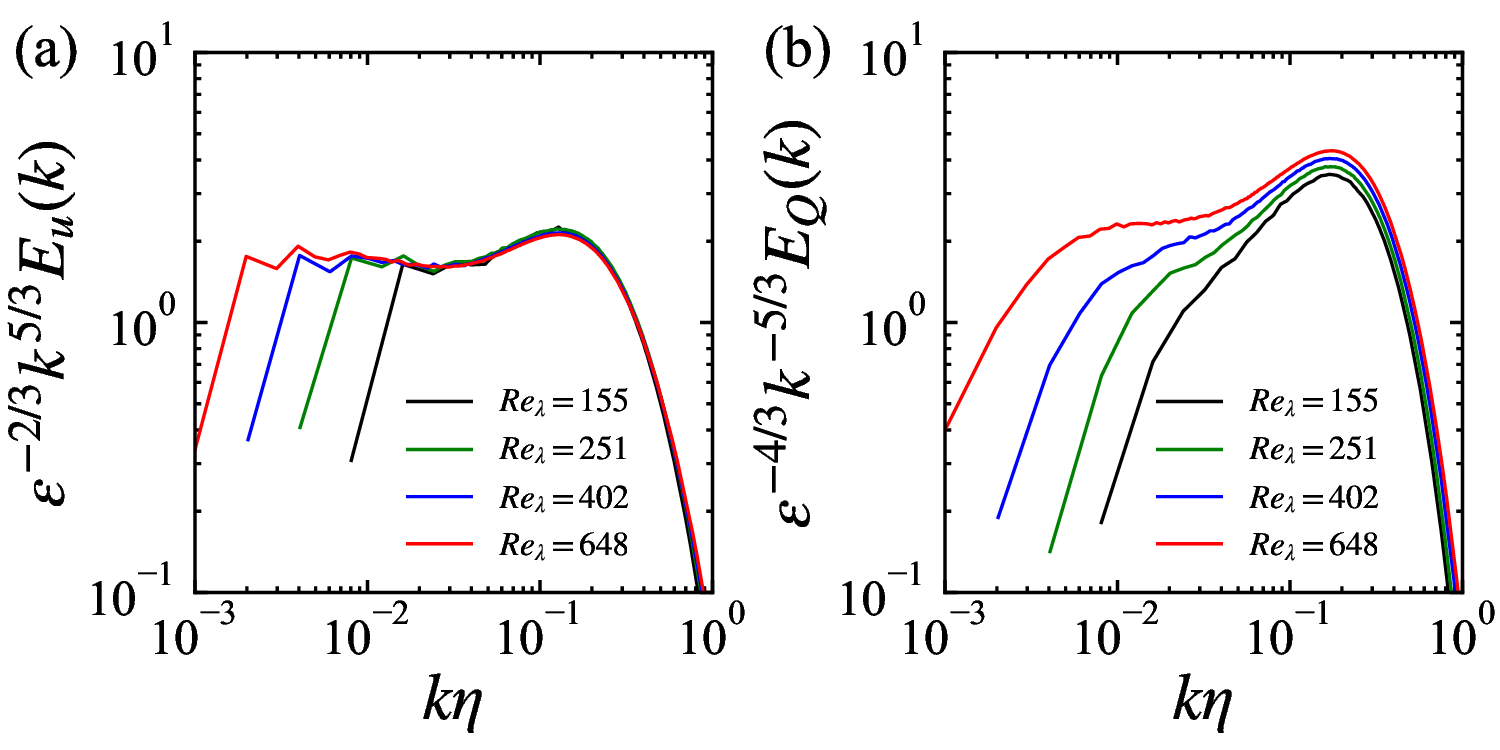}
    \caption{ 
    (a) Compensated kinetic energy spectra and (b) compensated spectra of $Q$. 
    }
    \label{fig:energy_spect}
\end{figure}
Figure~\ref{fig:energy_spect}(a) shows kinetic energy spectra $E_u(k)$ for the different flows listed in Table~\ref{tab:DNS}. 
Here, $E_u(k) \equiv (1/2)\sum'_k |\widehat{\bm u}(\bm k)|^2$, where $\widehat{\cdot}$ denotes the Fourier transform, $\sum'_k \equiv (4\pi k^2/N_k)\sum_{k-1/2 \le |{\bm k}| < k+1/2} $, and $N_k$ is the number of $\bm k$ that 
satisfy $k-1/2 \le |{\bm k}| < k+1/2$.
In the largest Reynolds number case, $Re_{\lambda} = 648$, the compensated spectrum is nearly flat for about one decade in $k\eta$, indicating that the spectrum is close to the scaling $k^{-5/3}$ \cite{K41}.
According to the preferential concentration mechanism \cite{Maxey,Squires&Eaton1990}, the clustering formation is affected by the vorticity and strain rate distributions represented by the second invariant $Q$ of the fluid velocity gradient tensor, 
where $Q \equiv [{\bm \Omega}{\bm :}{\bm \Omega} - {\bf S}{\bm :}{\bf S}]/2$ 
with ${\bm \Omega} \equiv [\nabla{\bm u} - (\nabla{\bm u})^T]/2$ 
and ${\bf S} \equiv [\nabla{\bm u} + (\nabla{\bm u})^T]/2$. 
$Q$ satisfies the relationship $2Q=\nabla^2 {\cal P}$.
The superscript $T$ denotes the transposed.
Positive and negative $Q$ indicate rotation- and strain-dominated regions, respectively.
According to dimensional analysis \cite{MY1975,GF2001,Ishihara2003} following K41, 
the spectrum 
\revision{of $Q$, defined as $E_Q(k) \equiv \sum'_k |\widehat{Q}(\bm k)|^2$,}
has a $k^{5/3}$ scaling in the turbulence inertial subrange at sufficiently high Reynolds number because $E_Q(k)$ is equivalent to $(1/4)k^4E_{\cal P}(k)$, where $E_{\cal P}(k) \equiv \sum'_k |\widehat{\mathcal P}(\bm k)|^2$ is the pressure spectrum.
\revision{Figure~\ref{fig:energy_spect}(b) shows the spectra $E_Q(k)$.}
This figure confirms that, for $Re_\lambda=648$, the compensated spectrum is almost flat for $0.008 \lesssim k\eta \lesssim 0.03$, 
implying that $E_Q(k)$ well obeys the prediction of the dimensional analysis. 
\revision{For $Re_\lambda=648$, the forcing is imposed in $|{\bm k}|\eta < 2.48 \times 10^{-3}$, which is smaller than the above wave number range.}

\begin{figure} 
    \centering
    \includegraphics[width=1.0\linewidth]{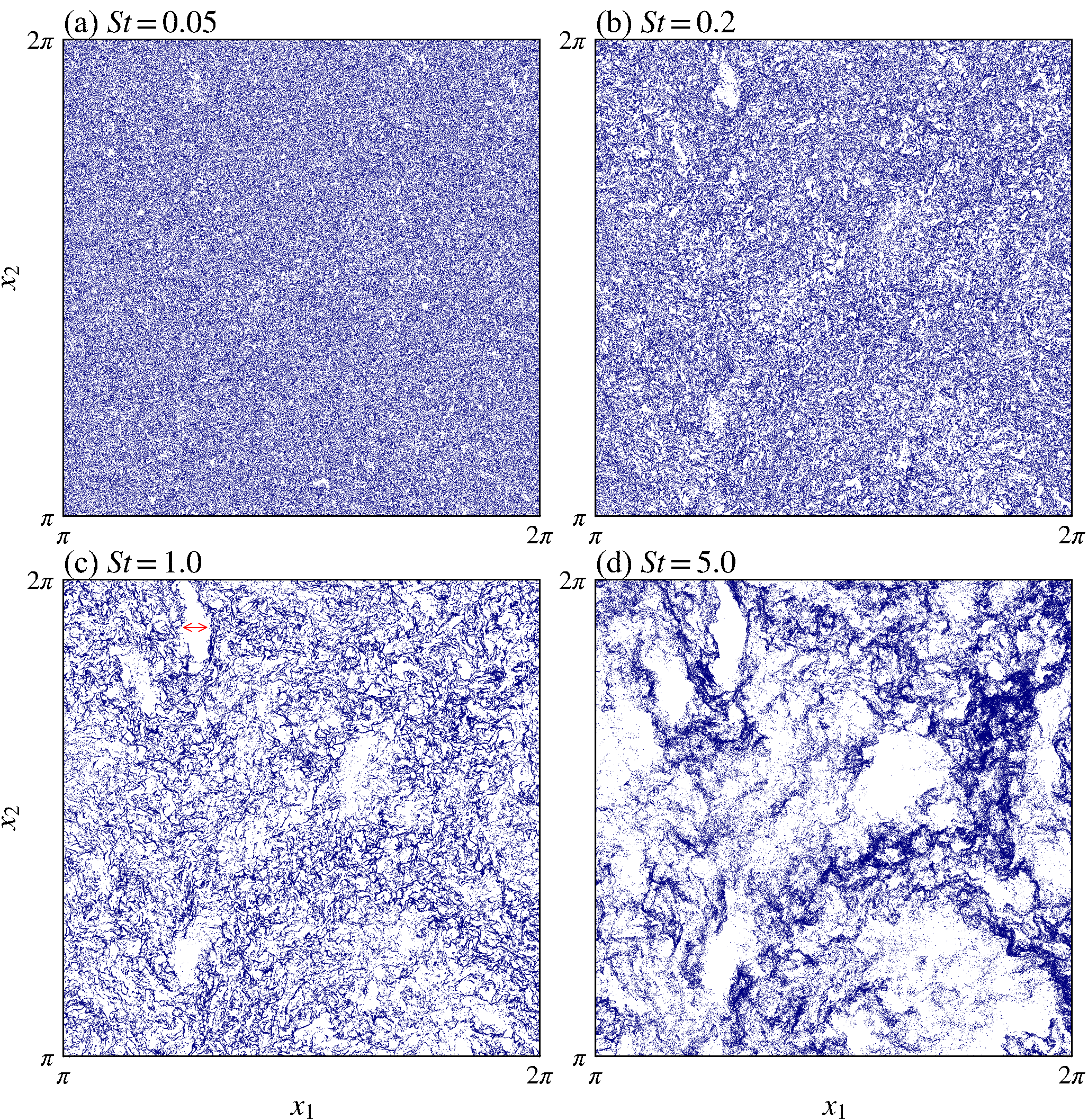}
    \caption{Spatial distributions of the particles for (a) $St=0.05$ (b) 0.2, (c) 1.0 and (d) 5.0 at $t=11T_0$ in the ranges of $\pi \le x_1 \le 2\pi$, $\pi \le x_2 \le 2\pi$, and $0 \le x_3 \le 4\eta$ for $Re_\lambda=648$. 
    The red arrow in (c) indicates the length $200\eta$.
    }
    \label{fig:pcl_2D}
\end{figure}

Spatial distributions of particles obtained by the DNS 
for $Re_\lambda=648$ are displayed in Fig.~\ref{fig:pcl_2D}. For $St=0.05$ and 0.2, small voids distribute intermittently, whereas for $St=1.0$ and 5.0, particle 
non-uniformity is significant even for scales larger than $200\eta$. 

We consider the particle number density field $n(\bm x,t)$ in the continuous setting \cite{Maxey} to define the particle number density spectrum $E_n(k) \equiv \sum'_k |\widehat{n}(\bm k)|^2$. 
To compute the spectrum, the discrete particle position data are converted into number density field data on $N_g^3$ equidistant grid points with the histogram method.
The number density field is normalized such that the mean value yields $\langle n \rangle = 1$.
The discrete nature of particles causes Poisson noise, and the conversion to the field data causes the suppression for large wave numbers \cite{SaitoGotoh2018}. 
These effects have been removed from the obtained spectra 
\revision{(see the SM \cite{SM})}.



\begin{figure}[tb!]
    \centering
    \includegraphics[width=8.6cm]{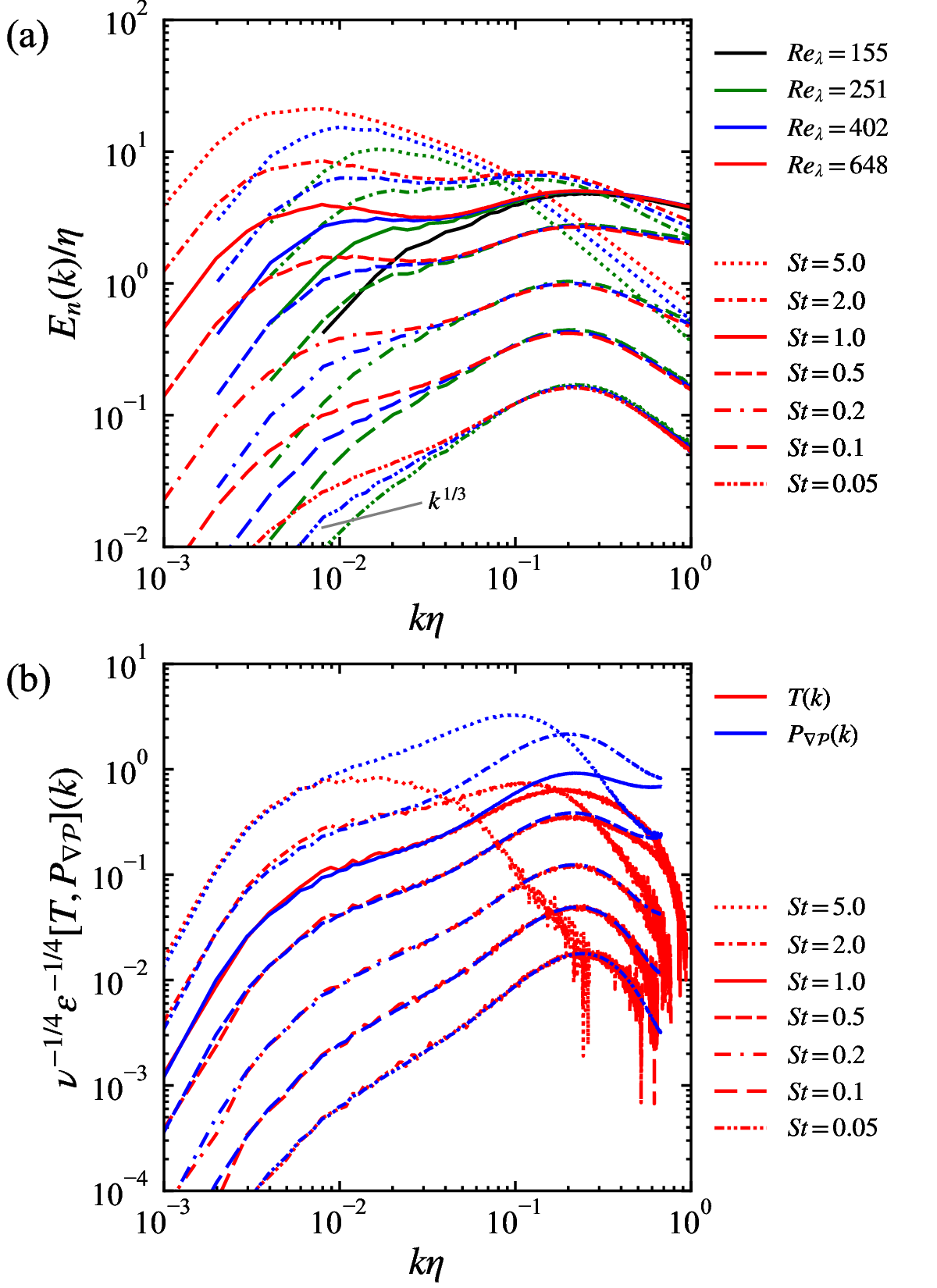}
    \caption{(a) Particle number density spectra $E_n(k)$ for $Re_\lambda=155$, 251, 402 and 648, and (b) transfer spectra $T(k)$ for $Re_\lambda=648$ with the approximated production spectra $P_{\nabla \cal P}(k)$.
    \revision{The number density spectra compensated by the scaling $k^{1/3}$ are provided in the SM \cite{SM}.}
    }
\label{fig:Enp_T_spect_St}
\end{figure}

Figure~\ref{fig:Enp_T_spect_St}(a) shows $Re_\lambda$ dependence of the number density spectra $E_n(k)$ for different $St$ values.
For $St=1.0$, each spectrum for $Re_\lambda > 200$ shows a peak around $k\eta \approx 0.2$.
The peak wave number of the bump near the dissipation range is similar for $St \le 1$.
These observations are consistent with the results in Ref.~\cite{Matsuda2014}. 
For $k\eta \lesssim 0.03$, the spectrum values increase as $Re_\lambda$ increases for all $St$ cases, 
and each spectrum for $St \ge 0.5$ clearly displays a bump for $Re_\lambda = 648$.
For $St=5.0$, only the bump for $k\eta \lesssim 0.03$ exists, and the peak near the dissipation range disappears.
Note that we confirmed that the existence of the bump for $k\eta \lesssim 0.03$ is robust against the external forcing schemes \cite{Onishi2011,YA2007} 
(See the SM \cite{SM} and Ref.~\cite{MatsudaTSFP12} therein).
We can also observe that the slope of the spectra in the inertial subrange ($0.008 \lesssim k\eta \lesssim 0.03$) is dependent on $St$ in contrast to the prediction by Ariki {\it et al.}~\cite{Ariki2018}, in which $E_n(k) \propto k^{1/3}$.
The slope for $St=0.05$ and $Re_\lambda=648$ is close to $1/3$.  
However, the negative slopes for $St \ge 0.5$ are obviously different from the prediction.

\revision{To examine the clustering mechanism in the inertial subrange, we consider the particle velocity 
field ${\bm v}({\bm x},t)$ in the continuous setting 
based on the approximation by Maxey \cite{Maxey}, i.e.,
\begin{equation}
    {\bm v} = {\bm u} -\tau_p {\bm a}_L +{\cal O}(\tau_p^2)
    \label{eq:v_Maxey},
\end{equation}
where ${\bm a}_L \equiv \partial{\bm u}/\partial t + {\bm u}{\bm \cdot}\nabla{\bm u}$, which is the Lagrangian acceleration of fluid.
Then, the conservation equation of $n$  
reads $\partial n/\partial t + \nabla {\bm \cdot} (n{\bm v}) = 0$. 
This approximation \revision{is} 
valid at least for a coarse-grained field of ${\bm v}_p$ for a spatial scale $r$
satisfying $\tau_p \ll \tau_r$, where 
$\tau_r$ is the time scale of the fluid flow for $r$. 
Note that Maxey \cite{Maxey} considered the Lagrangian transport of $n$ along the inertial particle trajectory,
$\partial n/\partial t + {\bm v} {\bm \cdot} \nabla n = -n \nabla {\bm \cdot} {\bm v}$, 
and applied Eq.~(\ref{eq:v_Maxey}) to the particle velocity divergence, yielding $\nabla {\bm \cdot} {\bm v} = 2\tau_p Q + {\cal O}(\tau_p^2)$.
The preferential concentration mechanism is explained by this approximation: 
the inertial particles preferentially concentrate in flow regions where $Q$ is small or negative.
}
\revision{The conservation equation} can be rewritten in the form of Lagrangian transport of $n$ along the fluid particle trajectory, 
\begin{equation}
    \frac{\partial n}{\partial t} + {\bm u} {\bm \cdot} \nabla n = - \nabla {\bm \cdot} \left( n{\bm v} - n{\bm u} \right) 
    \label{eq:n_conservation_vrel},
\end{equation}
in which the production term on the right-hand side represents the effect of particle drift velocity. 
Based on the advection term, $A({\bm x}) \equiv {\bm u} {\bm \cdot} \nabla n$, in Eq.~(\ref{eq:n_conservation_vrel}), the transfer spectrum for the number density is defined as $T(k) \equiv 2 \sum'_k \mathfrak{R}[ \widehat{A}(\bm k) \widehat{n}^*(\bm k) ]$\revision{, which represents the scale-to-scale transfer of number density fluctuation}. 
Here, $\mathfrak{R}[\cdot]$ denotes the real part and the asterisk the complex conjugate.
Figure~\ref{fig:Enp_T_spect_St}(b) shows the transfer spectra $T(k)$ computed for each $St$ for $Re_\lambda = 648$.
For all the Stokes numbers, $T(k)$ exhibits positive values in the inertial range, indicating that the forward transfer of particle number density fluctuation (from large scales to small scales) is dominant. 
Thus, the bump of $E_n(k)$ for $k\eta \lesssim 0.03$ is due to pronounced clustering production in the inertial range and not due to backward transfer from the dissipation range.

We analyze the clustering production in the inertial subrange based on the production term, $B({\bm x}) \equiv -\nabla{\bm \cdot}(n{\bm v} - n{\bm u})\revision{= \tau_p \nabla{\bm \cdot}(n{\bm a}_L) + {\cal O}(\tau_p^2)}$, in Eq.~(\ref{eq:n_conservation_vrel})\revision{.} 
When considering the Fourier transform $\widehat{B}(\bm k)$ in the inertial subrange,
the effect of viscosity and forcing terms on ${\bm a}_L$ is negligibly small according to the K41 phenomenology.
Therefore, we can assume that only the pressure gradient term in the N--S equation affects the inertial particle drift, i.e., 
$\widehat{B}(\bm k) \approx \widehat{B_{\nabla {\cal P}}}(\bm k)$, where $B_{\nabla {\cal P}}(\bm x) \equiv - \tau_p \nabla{\bm \cdot}(n\nabla {\cal P})$. 
Then we can define the approximated production spectrum $P_{\nabla{\cal P}}(k) \equiv 2 \sum'_k \mathfrak{R}[ \widehat{B_{\nabla{\cal P}}}(\bm k) \widehat{n}^*(\bm k) ]$.
Note that the exact production spectrum is equivalent to $T(k)$ assuming statistical stationarity. 
Figure~\ref{fig:Enp_T_spect_St}(b) shows the approximated spectra $P_{\nabla{\cal P}}(k)$. 
For $St \le 2.0$, $P_{\nabla{\cal P}}(k)$ well agrees with $T(k)$ for $k\eta \lesssim 0.03$, including the inertial subrange. 
For $St=5.0$, $P_{\nabla{\cal P}}(k)$ and $T(k)$ show significant difference in the inertial subrange, 
indicating a breakdown of the approximation (\ref{eq:v_Maxey}) due to large $\tau_p$.
Therefore, we can conclude that $B_{\nabla {\cal P}}$ plays a dominant role for the clustering production for $k\eta \lesssim 0.03$ and $St \le 2.0$, as predicted above.

The approximated production term $B_{\nabla{\cal P}}$ represents the production due to the convergence of the particle inertial drift, and can be rewritten as $B_{\nabla{\cal P}} = B_Q + B_\theta$, where $B_Q \equiv -2\tau_p Q$ and $B_\theta \equiv -\tau_p \nabla{\bm \cdot}(\theta \nabla {\cal P})$ with $\theta \equiv n - 1$, where $\langle \theta \rangle =0$. 
The term $B_Q$ represents the preferential concentration mechanism, and its contribution can be dominant when the number density is nearly uniform. 
$B_{\nabla{\cal P}}$ guarantees the conservation of $n$ but $B_Q$ does not. Hence the residual term $B_\theta$ can be considered as the modification of the clustering production for the conservation of $n$.
The effect of $B_\theta$ can be significant reflecting the non-uniformity of $n$.

We examine the scale dependence of the number density, transfer and production spectra in the approximated balance equation $\partial E_n(k)/\partial t + T(k) = P_{\nabla{\cal P}}(k)$, using dimensional analysis.
Ariki {\it et al.}~\cite{Ariki2018} predicted a scale similarity solution $kE_n(k) \propto St_r^2$ using the approximation $B \approx B_Q$ for $\tau_p \ll \tau_r$. 
Here, $St_r \equiv \tau_p/\tau_r$ is the scale-dependent Stokes number \cite{Bec2007,Bragg2015,Ariki2018} \revision{with $\tau_r=\epsilon^{-1/3}k^{-2/3}$ for a scale $r=k^{-1}$}.
Following K41, the representative velocity and pressure at the scale $k$ are scaled by $\epsilon^{1/3} k^{-1/3}$ and $\epsilon^{2/3} k^{-2/3}$ in the inertial subrange, respectively \cite{K41,MY1975}.
Then, we can expect $k E_n(k)$, $\tau_r k T(k)$ and $\tau_r k P_{\nabla{\cal P}}(k)$ to be given by a function of $St_r$ even when the contribution of $B_\theta$ in $B_{\nabla{\cal P}} = B_Q + B_\theta$ is not negligible.

\begin{figure}[tb!]
    \centering
    \includegraphics[width=8.6cm]{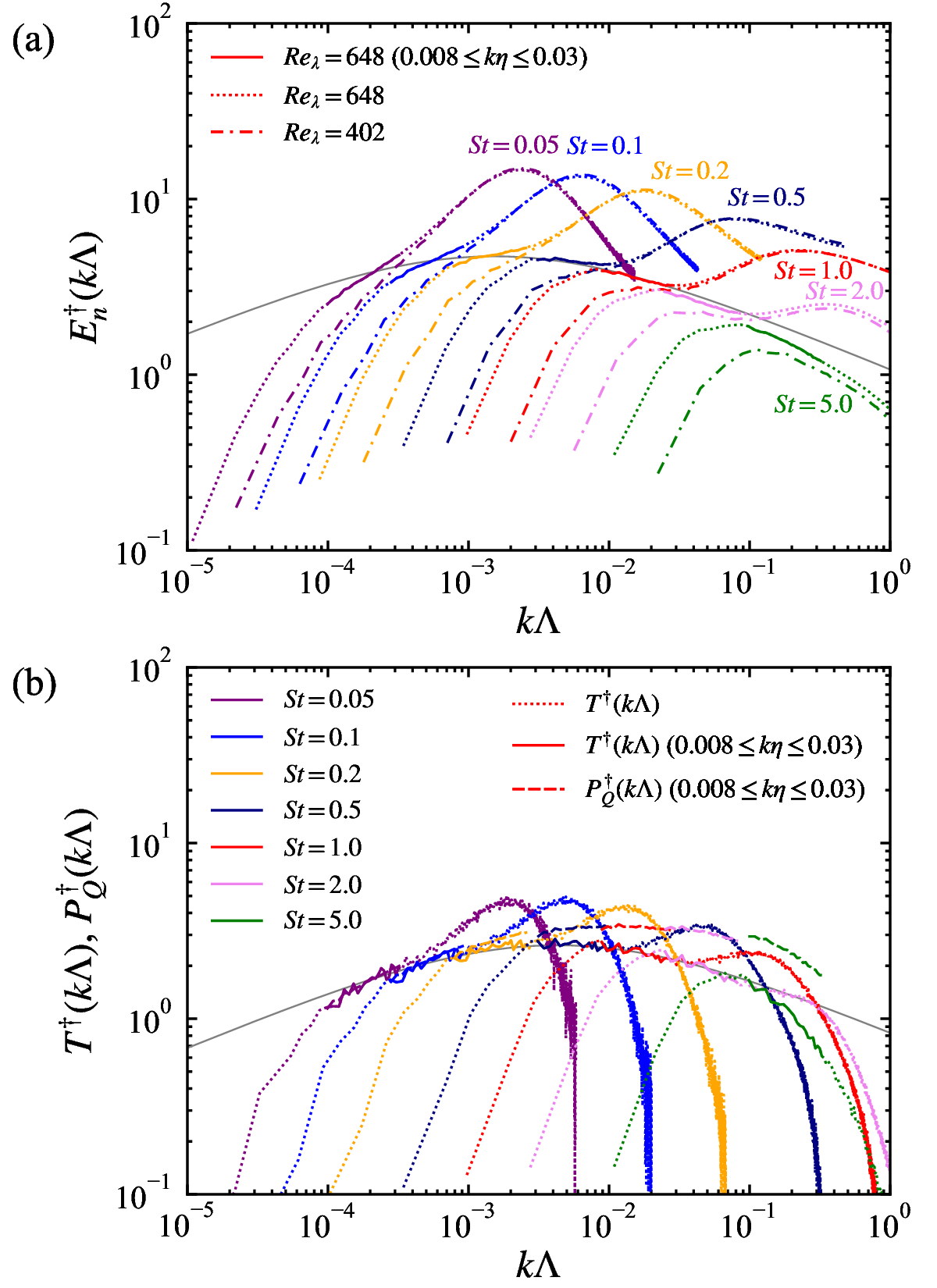}
    \caption{(a) Normalized number density spectra $E_n^\dag(k\Lambda)$ for $Re_\lambda=648$ and 402,  
    and (b) normalized transfer spectra $T^\dag(k\Lambda)$ for $Re_\lambda=648$. 
    The spectra in the inertial subrange ($0.008 \lesssim k\eta \lesssim 0.03$) are indicated by the solid lines. 
    The dashed lines in (b) are the normalized spectra $P_Q^\dag(k\Lambda)$ for the production due to $B_Q$ for $Re_\lambda=648$ only in the inertial subrange.
    The gray solid lines in (a) and (b) are approximated functions for the inertial subrange, 
    $E_n^\dag(k\Lambda)\approx 82.1 (k\Lambda)^{1/3} /(1 + 75.7 St_r)$ and 
    $T^\dag(k\Lambda)\approx 32.2 (k\Lambda)^{1/3}/(1 + 37.8 St_r)$, respectively, 
    where $k\Lambda = St_r^{3/2}$.
    }
\label{fig:Enp_T_spect_kLambda}
\end{figure}

This expectation is verified by normalizing the spectra by a characteristic length $\Lambda \equiv \tau_p^{3/2}\epsilon^{1/2}$ \cite{Ariki2018} \revision{so that}  
$k\Lambda = St_r^{3/2}$\revision{, i.e., $\tau_r \sim \tau_p$ for $r \sim \Lambda$}.
The normalized spectra $E_n^\dag(k\Lambda) \equiv E_n(k)/\Lambda$ and $T^\dag(k\Lambda) \equiv \tau_r T(k)/\Lambda$ are displayed in Fig.~\ref{fig:Enp_T_spect_kLambda}. 
The spectra $E_n^\dag(k\Lambda)$ in the inertial subrange, indicated by solid lines in Fig.~\ref{fig:Enp_T_spect_kLambda}(a), 
\revision{align} on the same curve,
which is a function of $k\Lambda$,  
for $Re_\lambda=648$ and $St=0.5$, 1.0 and 2.0.
The spectra for $St=0.05$, 0.1 and 0.2 lay only near the curve.
For the case of $Re_\lambda=402$, the spectra $E_n^\dag(k\Lambda)$ in the inertial subrange do not show the alignment on the same curve because $Q$ and $\mathcal P$ do not obey the scaling predicted by the dimensional analysis (See Fig.~\ref{fig:energy_spect}(b)).
In Fig.~\ref{fig:Enp_T_spect_kLambda}(b), the transfer spectra $T^\dag(k\Lambda)$ for $Re_\lambda=648$ align along a single curve in the inertial subrange more clearly for $0.05 \le St \le 2.0$. 
These results support our conjecture that the spectra are given by a function of $St_r$, whereas the slope of both spectra changes significantly depending on $k\Lambda$, i.e., $St_r$.
Therefore, the bump observed in $E_n(k)$ for $k\eta \lesssim 0.03$ for $St \ge 0.5$ is due to the $St_r$ dependence of $E_n^\dag(k\Lambda)$ in the inertial subrange, which shows a negative slope for $k\Lambda \gtrsim 10^{-3}$ ($St_r \gtrsim 10^{-2}$).
The normalized production spectra $P_Q^\dag(k\Lambda) \equiv \tau_r k P_Q(k)$, where $P_Q(k) \equiv 2 \sum'_k \mathfrak{R}[ \widehat{B_Q}(\bm k) \widehat{n}^*(\bm k)]$, i.e., the contribution of the preferential concentration, are also displayed in Fig.~\ref{fig:Enp_T_spect_kLambda}(b), only in the inertial subrange. 
A clear discrepancy between $T^\dag(k\Lambda)$ and $P_Q^\dag(k\Lambda)$ can be observed for $k\Lambda \gtrsim 10^{-3}$, meaning that the clustering production for such scales is reduced by the contribution of $B_\theta$, 
i.e., the modulation due to the $n$ conservation. 
The observed $St_r$ dependence of $E_n^\dag(k\Lambda)$ implies that the spatial structure of $n$ at each scale changes depending on $St_r$.
It has been confirmed in the SM \cite{SM}
that the skewness and flatness factors at each scale of $n$ well exhibit the scale dependence as a function of $St_r$.

If the $St_r$ dependence of $E_n^\dag(k\Lambda)$ holds even for higher $Re_\lambda$, then we conjecture that the $St_r$ dependence of the slope should be observed in the inertial subrange for $St \gtrsim 0.5$.
The Kolmogorov-like scaling predicted in Ref.~\cite{Ariki2018} could be observed only for smaller values of $St_r$, i.e., $St_r \ll 10^{-2}$, supposing such a scaling exists.
Assuming that the predicted scale similarity appears for $St_r \ll 1$, 
$E_n^\dag(k\Lambda)$ and $T^\dag(k\Lambda)$ can be approximated by functions.
A least-square fitting of a rational function of the form $c_1/(1 + c_2 St_r)$ with constants $c_1$ and $c_2$ to $(k\Lambda)^{-1/3} E_n^\dag(k\Lambda)$ and $(k\Lambda)^{-1/3} T^\dag(k\Lambda)$ in the inertial subrange results in the gray solid curves in Fig.~\ref{fig:Enp_T_spect_kLambda}.

\revision{
Experimental studies are also crucial to verify the scale dependence of the clustering for high Reynolds numbers.  
In Ref.~\cite{Saw2012}, a plateau in the RDFs, referred to as a `shoulder region',
was reported for $r/\eta \ge 50$ based on wind tunnel experiments for $Re_\lambda$ up to 800. 
However, the authors stated that it was attributed to large-scale inhomogeneity caused by the experimental setup.
Reference \cite{Petersen2019} reported that the clustering was observed up to the scales $300\eta$--$400\eta$ in experiments for $Re_\lambda=500$.
}
\revision{
In practical situations, e.g., laboratory experiments or cloud droplets and aerosols in the atmosphere, the gravitational settling further influences the clustering (e.g., Refs.~\cite{Bec2014,Gustavsson2014,Ireland2016b,Petersen2019}). 
Under the presence of gravity, the particle motion can be decoupled from the carrier turbulent flow. 
In the inertial subrange, the particle clustering would be modulated when the terminal velocity $v_T$ is larger than the flow velocity scale $u_r = \epsilon^{1/3} k^{-1/3}$ for a scale $k$.
Therefore, for large scales that satisfy $v_T \ll u_r$, the settling effect could remain negligibly small as discussed in Ref.~\cite{Lu2010}.  
}

In this work, the pronounced inertial particle clustering has been discovered in the inertial subrange, in addition to the clustering in the near dissipation range. 
The obtained number density spectra well obey a function of $St_r$ in the inertial subrange, showing a bump around $k\Lambda \approx 10^{-3}$.
The clustering production is dominated by the preferential concentration mechanism for $k\Lambda \lesssim 10^{-3}$, whereas it is suppressed by the modulation due to the $n$ conservation for $10^{-3} \lesssim k\Lambda \lesssim  10^{-1}$.
The discovered inertial particle clustering in the inertial subrange could have an impact on the subgrid-scale modeling for large-eddy simulations.
The clustering of cloud droplets in the inertial subrange may also cause temperature and moisture fluctuations, which affect the condensation and evaporation of droplets in the raindrop formation process. 
\revision{Including the influence of gravitational setting, the spatial correlation for different Stokes numbers and turbulent mixing of clear air, in future work is important to understand and to model the droplet behavior in cloud turbulence.}

\bigskip
\begin{acknowledgments}
KM acknowledges financial support from JSPS KAKENHI Grant Numbers JP20K04298 and JP23K03686.
The direct numerical simulations presented here were conducted on the Earth Simulator supercomputer system operated by JAMSTEC.
\end{acknowledgments}

\pagebreak
\onecolumngrid
\begin{center}
\textbf{\large 
Supplemental Material: 
Heavy Particle Clustering in Inertial Subrange of High--Reynolds Number Turbulence
}
\end{center}
\setcounter{equation}{0}
\setcounter{figure}{0}
\setcounter{table}{0}
\setcounter{page}{1}
\makeatletter
\renewcommand{\theequation}{S\arabic{equation}}
\renewcommand{\thefigure}{S\arabic{figure}}
\renewcommand{\thetable}{S\Roman{table}}
\renewcommand{\bibnumfmt}[1]{[S#1]}
\renewcommand{\citenumfont}[1]{S#1}

\twocolumngrid

\section{Sensitivity to external forcing schemes}

In this work, the large-scale random forcing (RF) scheme was employed following Ref.~\cite{YA2007} for the external solenoidal forcing ${\bm f}({\bm x},t)$ to obtain statistically stationary turbulence in the DNS.
Here, we describe the detail of the scheme and assess the sensitivity of the spectra to the forcing scheme.
The large-scale RF scheme calculates the forcing $\bm f$ 
in Fourier space based on 
\begin{eqnarray}
\widehat{\bm f}({\bm k},0) &=& \sqrt{F} {\bm R}({\bm k},0), {\rm and} \\
\widehat{\bm f}({\bm k},t+\Delta t) &=& \gamma \widehat{\bm f}({\bm k},t)+\zeta \sqrt{\Delta t} {\bm R}({\bm k},t),
\end{eqnarray}
where $\gamma \equiv e^{-\Delta t/T_f}$, $\zeta \equiv \sqrt{2F/T_f}$, and ${\bm R}({\bm k},t)$ is a vector based on Gaussian random variables 
with ${\bm k}$ being the wave vector. 
$T_f$ and $F$ are the expected values of the correlation time and intensity of $\bm f$, respectively. 
Here, we used $T_f = T_0$ and $F=4.8 \times 10^{-3}$. 
\revision{The forcing was applied for large scales $|{\bm k}|<2.5$. The corresponding $|{\bm k}|\eta$ range is $|{\bm k}|\eta < 2.00 \times 10^{-2}$, $1.01 \times 10^{-2}$, $5.05 \times 10^{-3}$ and $2.48 \times 10^{-3}$ for $Re_\lambda=155$, 251, 402 and 648, respectively.}

To assess the sensitivity to the forcing scheme, we have also employed the large-scale linear forcing (LF) scheme in Ref.~\cite{Onishi2011} for comparison with the large-scale RF scheme.
The large-scale LF scheme determined the forcing $\bm f$ such that $\widehat{\bm f}({\bm k},t)$ is proportional to the Fourier component of velocity $\widehat{\bm u}({\bm k},t)$ for $|{\bm k}|<2.5$. 
The DNS with the LF scheme was performed for the flows listed in Table~\ref{tab:DNS_LF}.
\revision{The $|{\bm k}|\eta$ range for the LF is $|{\bm k}|\eta < 1.97 \times 10^{-2}$, $1.04 \times 10^{-2}$ and $5.23 \times 10^{-3}$ for $Re_\lambda=204$, 328 and 531, respectively.}
For the case of the RF, the forcing $\bm f$ changes gradually by the random process, and the strong vortices could decay at the time scale of the forcing correlation time $T_f (=T_0)$. For the case of the LF, the forcing $\bm f$ tends to retain the large-scale velocity vector, and large-scale vortices could survive longer than the eddy-turnover time ($\sim T_0$). 
\revision{Note that for both RF and LF, ${\bm f}$ is divergence free.}

\begin{table}[b!]
\begin{center}
\begin{tabular}{ccccccc}
\hline\hline
 $N_g$ & $N_p$ & $Re$ & $Re_\lambda$ & $u'$ & $\epsilon$ & $k_{\mathrm{max}}\eta$ \\
\hline
 512  & 1.5 $\times 10^7$ & 909   & 204 & 1.01 & 0.343 & 2.02 \\
 1024 & 5.0 $\times 10^7$ & 2220  & 328 & 1.00 & 0.312 & 2.12 \\
 2048 & 4.0 $\times 10^8$ & 5595  & 531 & 1.00 & 0.298 & 2.14 \\
\hline\hline
\end{tabular}
\end{center}
\caption{DNS parameters and statistics of turbulence computed by using the large-scale linear forcing scheme 
(as in Table I in the main manuscript).
}
\label{tab:DNS_LF}
\end{table}

\begin{figure}[tb!]
    \centering
    \includegraphics[width=8.6cm]{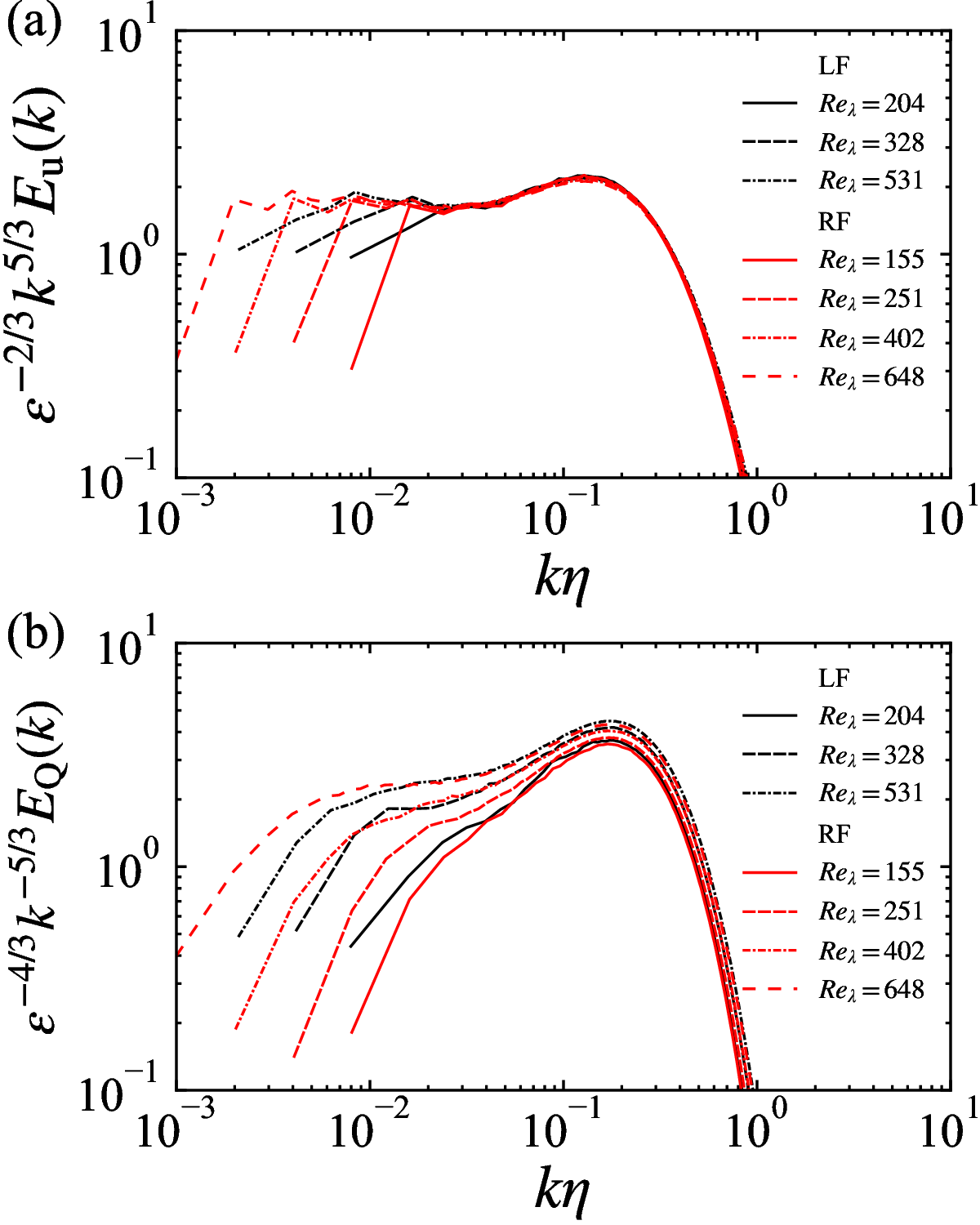}
    \caption{ 
    (a) Compensated kinetic energy spectra and (b) compensated spectra of $Q$ for the LF and RF cases.
    }
    \label{fig:energy_spect_LFRF}
\end{figure}

The kinetic energy spectra $E_u(k)$ and the $Q$ spectra $E_Q(k)$ obtained by using the different forcing schemes are compared in Fig.~\ref{fig:energy_spect_LFRF}. 
For both forcing schemes, the scaling of $E_u(k)$ is close to the $k^{-5/3}$ scaling. 
It is also observed that, for $Re_\lambda > 500$, the compensated spectra of $E_Q(k)$ show a range where the spectrum well obeys the $k^{5/3}$ scaling 
expected for the inertial subrange at sufficiently high Reynolds number \cite{GF2001,Ishihara2003}. 

\begin{figure}[tb!]
    \centering
    \includegraphics[width=8.6cm]{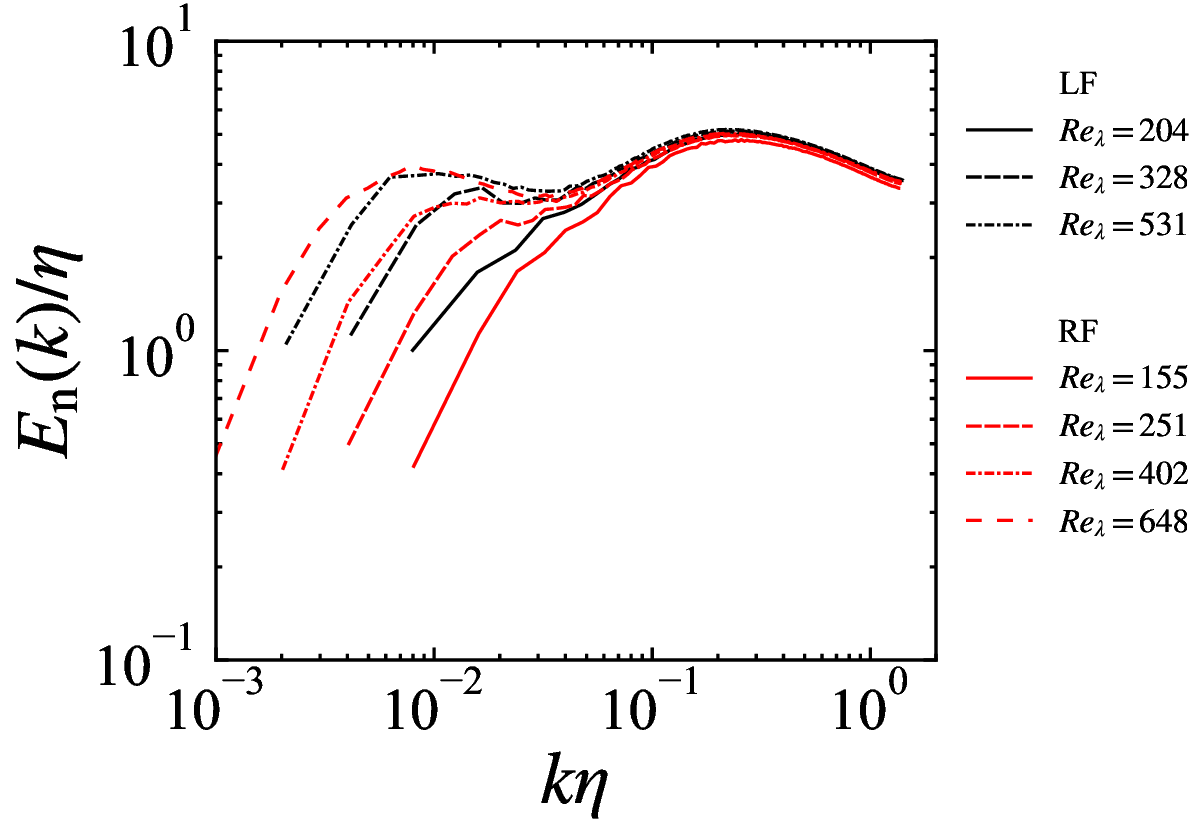}
    \caption{Particle number density spectra for $St=1.0$
    using the LF and RF schemes.
    }
\label{fig:Enp_spect_LFRF}
\end{figure}

The number density spectra $E_n(k)$ using the LF and RF are compared in Fig.~\ref{fig:Enp_spect_LFRF}.
For both cases of forcing schemes, the spectrum shows a bump for $k\eta \lesssim 0.03$ for $Re_\lambda > 300$, 
and the bump becomes higher as $Re_\lambda$ increases. 
Ref.~\cite{Matsuda2014} also showed a similar $Re_\lambda$ dependence of the number density spectrum 
for $k\eta \lesssim 0.03$ by increasing $Re_\lambda$ up to 531 and using the large-scale LF scheme.
These results indicate that the existence of the bump for $k\eta \lesssim 0.03$ is 
robust against the forcing schemes used for maintaining the turbulent flow \cite{MatsudaTSFP12}.

\revision{
\section{Computation of number density spectrum}
}

\revision{
Discrete nature of particles can cause Poisson noise in the number density spectrum $E_n(k)$ \cite{Matsuda2014}.
In addition, conversion to field data can cause suppression of the spectrum for large wave numbers \cite{SaitoGotoh2018}. 
In the following we explain the computational method of the spectra removing those effects.
}

\revision{
We consider the number density field of the discrete particle positions using the Dirac delta function $\delta({\bm x})$, 
\begin{equation}
    n_{\delta}({\bm x},t) \equiv \frac{1}{n_0} \, \sum_{m=1}^{N_p} \delta \left({\bm x} - {\bm x}_{p,m}(t) \right)
    \label{eq:n_delta},
\end{equation}
where the subscript $m$ denotes the identification number of the particle, and $n_0$ is the scaling factor.
We use $n_0 = N_p/(2\pi)^3$ so that the mean value yields $\langle n_\delta \rangle = 1$.
}

\revision{
The discrete particle position data can be converted into number density field data on equidistant grid points based on the kernel density estimation, 
which is defined as
\begin{equation}
    n_K({\bm x},t) \equiv \int_{\mathbb{T}^3} K_h({\bm x} - {\bm x}')n_\delta({\bm x}',t) d{\bm x}' 
    \label{eq:n_K},
\end{equation}
where $\mathbb{T} = 2 \pi \mathbb{R} / {\mathbb{Z}}$, and $K_h({\bm x})$ is a sampling kernel centered at $\bm x$ for the scale $h$.
In this work, we use the histogram method, which corresponds to the zeroth-order kernel density estimation:
$K_h({\bm x})$ is a piecewise constant function defined as $K_h({\bm x})=1/h^3$ for $-h/2 \le x_i < h/2$ ($i=1, 2, 3$), 
while $K_h({\bm x})=0$ otherwise.
When the computational domain was discretized into an array of $N_g^3$ equally sized boxes, $h$ corresponds to the box width, i.e., $h=2\pi/N_g$.
Note that Eq.~(\ref{eq:n_K}) satisfies $\langle n_K \rangle = 1$.
}

\revision{
The number density spectrum $E_n(k)$ is defined by using the number density $n({\bm x})$ in the continuous setting, and the spectrum is computed by $E_n(k)=\sum'_k \widehat{\Phi}({\bm k})$, where $\widehat{\Phi}({\bm k}) \equiv |\widehat{n}(\bm k)|^2$ is the spectral density function for $n({\bm x})$.
However, $\widehat{\Phi}({\bm k})$ cannot be directly calculated from the Fourier transform of $n_\delta({\bm x})$ and $n_K({\bm x})$ due to Poisson noise and the sampling kernel.
The source of Poisson noise can be explained by using the spatial correlation function for $n_\delta({\bm x})$, defined as $\Phi_\delta({\bm r})\equiv \langle n_\delta({\bm x}+{\bm r}) n_\delta({\bm x}) \rangle$. This is related to the spatial correlation function $\Phi({\bm r}) \equiv \langle n({\bm x}+{\bm r}) n({\bm x}) \rangle$ via
\begin{equation}
    \Phi_\delta(\bm r) = \frac{1}{n_0}\delta(\bm r) + \Phi(\bm r) 
    \label{eq:Phi_delta_decomp2}.
\end{equation}
The first term on the right hand side comes from the correlation between identical particles 
and results in Poisson noise.
The influence of this term does not disappear even when using the kernel density estimation. The spatial correlation function $\Phi_K(\bm r) \equiv \langle n_K({\bm x}+{\bm r}) n_K({\bm x}) \rangle$ is given by  
\begin{equation}
\Phi_K(\bm r) = \frac{1}{n_0}K_h^2(\bm r) + \int_{\mathbb{T}^3} K_h^2(\bm r - \bm r') \Phi(\bm r') d\bm r' 
    \label{eq:Phi_K_decomp}.
\end{equation}
The Fourier transform of Eq.~(\ref{eq:Phi_delta_decomp2}) yields 
$\widehat{\Phi_\delta}(\bm k) = 1/N_p + \widehat{\Phi}(\bm k)$.
Obviously the first term causes Poisson noise in the spectrum. 
A Fourier transform method for discrete particle distribution was proposed in Ref.~\cite{Matsuda2014} removing Poisson noise based on this relationship.
When using the fast Fourier transform, we have to consider $\Phi_K(\bm r)$ instead, as the particle number density is given by the histogram. 
By applying the Fourier transform to Eq.~(\ref{eq:Phi_K_decomp}), 
the spectral density function $\widehat{\Phi_K}(\bm k) \equiv |\widehat{n_K}(\bm k)|^2$ is given by
\begin{equation}
\widehat{\Phi_K}(\bm k) 
    = \frac{1}{n_0}\widehat{K_h^2}(\bm k) 
    + (2\pi)^6 \widehat{K_h}(\bm k)^2 \widehat{\Phi}(\bm k) 
    \label{eq:Phi_K_hat},
\end{equation}
where $\widehat{K_h^2}(\bm k)$ is the Fourier coefficient of $K_h^2(\bm x)$, which is defined as $K_h^2(\bm x)=\int_{\mathbb{T}^3} K_h({\bm x} - {\bm x}')K_h({\bm x}') d{\bm x}'$. The Fourier transform of the convolution gives the product, i.e., $\widehat{K^2_h} = \widehat{K_h}^2$. 
The Fourier transform yields $(2\pi)^3\widehat{K_h}(\bm k) =  {\rm sinc}(k_1 h/2) {\rm sinc}(k_2 h/2) {\rm sinc}(k_3 h/2)$, where $\bm k = (k_1, k_2, k_3)$ and ${\rm sinc}(x) = \sin(x)/x$. 
It should be noted that the discrete Fourier transform of $K_h^2({\bm x})$ gives accurate estimate for Poisson noise because that can include the aliasing effect on Poisson noise. 
Since we use the piecewise constant function for $K_h({\bm x})$, 
the discrete Fourier transform of $K_h({\bm x})$ yields $\widehat{K_h^2}(\bm k)=(2\pi)^{-3}$. 
For the second term, the discrete Fourier transform does not yield a simpler formula. 
Thus, the spectral density function $\widehat{\Phi}({\bm k})$ is given by 
\begin{equation}
 \widehat{\Phi}({\bm k}) = \frac{1}{\kappa(\bm k)^2} \left[ \widehat{\Phi_K}(\bm k) - \frac{1}{N_p} \right]
    \label{eq:Phi_comput},
\end{equation}
where $\kappa({\bm k}) \equiv (2\pi)^3 \widehat{K_h}(\bm k)$. 
}

\bigskip

\revision{
\section{Compensated number density spectrum}
}

\revision{
The compensated number density spectra based on the prediction of Ariki {\it et al}. \cite{Ariki2018} are displayed in Figure~\ref{fig:Enp_spect_compensated}.
For $Re_\lambda=648$, the spectra are almost flat for $St=0.05$ and 0.1 in the inertial subrange, i.e., $0.008 \lesssim k\eta \lesssim 0.03$.
For larger $St$, the slopes are negative in the inertial subrange. 
For smaller $Re_\lambda$, the spectra do not show a flat plateau even for small $St$. 
\begin{figure}[tb!]
    \centering
    \includegraphics[width=8.6cm]{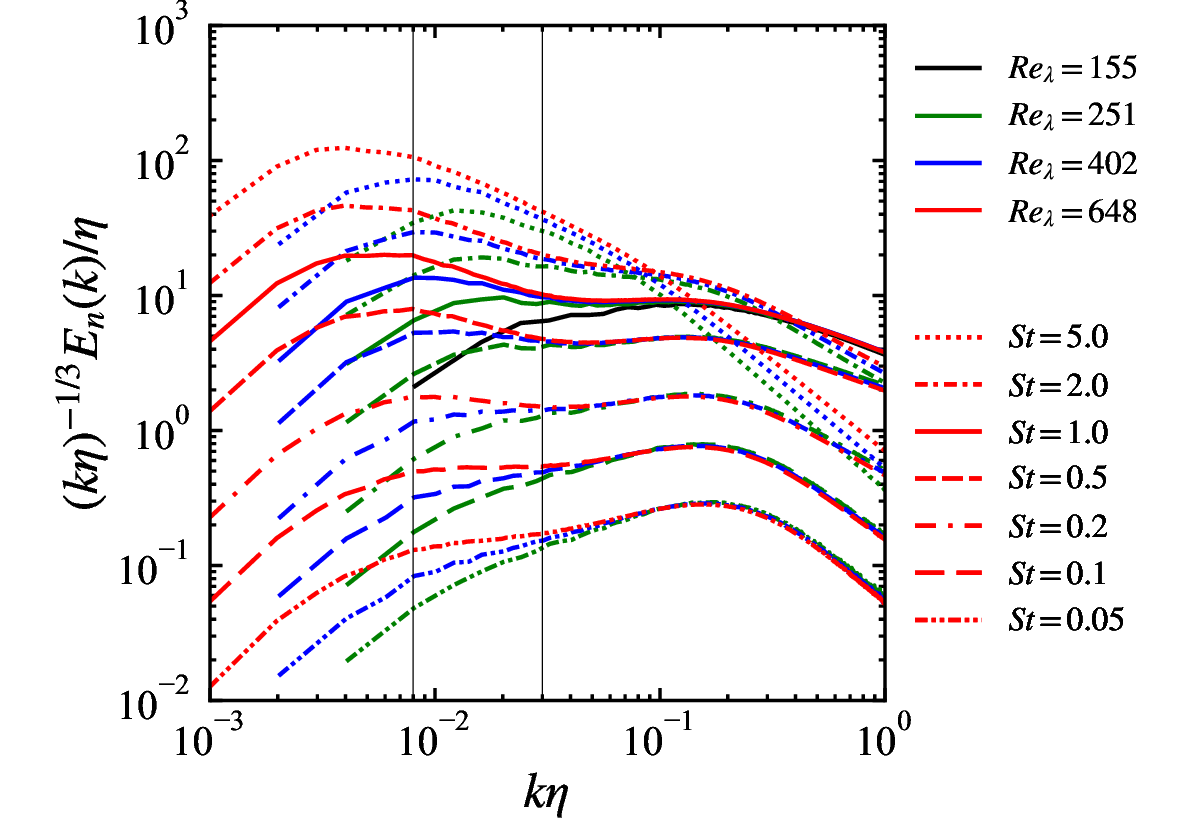}
    \caption{Particle number density spectra compensated by the scaling $k^{1/3}$.
    }
\label{fig:Enp_spect_compensated}
\end{figure}
}

\bigskip
\bigskip

\section{Scale dependence of skewness and flatness values}

\begin{figure}[h!]
    \centering
    \vspace{5mm}
    \includegraphics[width=8.6cm]{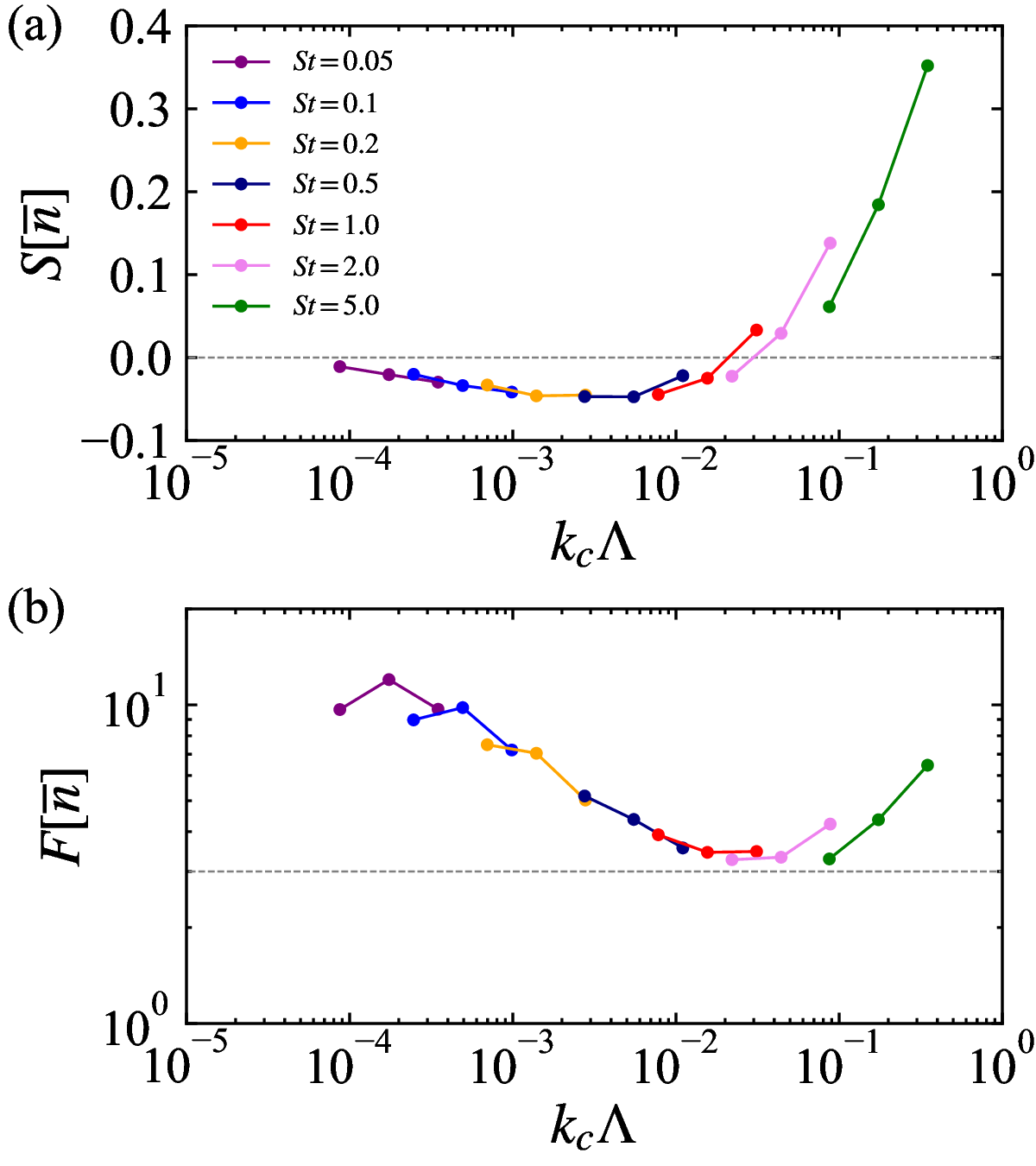}
    \caption{(a) Skewness and (b) flatness of band-pass filtered field $\overline{n}$ for $Re_\lambda=648$. 
    The values for $k_c\eta \approx 7.92\times10^{-3}$, $1.58\times10^{-2}$ and $3.17\times10^{-2}$ are plotted.
    }
    \label{fig:Skewness_Flatness}
\end{figure}

Scale dependence of the spatial structure of $n$ is examined by using the skewness and flatness factors of band-pass filtered number density fields $\overline{n}({\bm x}| k_c)$ in physical space.
Here, we use the difference of two Gaussian filters defined as $\widehat{G}({\bm k}|k_c)-\widehat{G}({\bm k}|k_c/2)$ in Fourier space, where $\widehat{G}({\bm k}|k_c) = \exp\left\{-\pi^2 |{\bm k}|^2/(24k_c^2)\right\}$, and $k_c$ is the cutoff wave number. 
The skewness and flatness factors are time-averaged for the particle position data at the ten time instants.
Figure~\ref{fig:Skewness_Flatness} shows the skewness and flatness factors, 
$S[\overline{n}] \equiv \langle \overline{n}^3 \rangle/\langle \overline{n}^2 \rangle^{3/2}$ and $F[\overline{n}] \equiv \langle \overline{n}^4 \rangle/\langle \overline{n}^2 \rangle^2$, respectively. 
These factors become $S[\overline{n}]=0$ and $F[\overline{n}]=3$ when a probability density function of $\overline{n}$ is given by a Gaussian distribution.
We can observe that the skewness values coincide in a single curve particularly for $k\Lambda \lesssim 10^{-2}$, and the flatness values also align around a single curve. 
Therefore, the spatial structure of $\overline{n}$ changes gradually along with $St_r$.
The skewness exhibits negative values for $k\Lambda \lesssim 0.02$, whereas it becomes positive for larger $k\Lambda$. This suggests a transition from void-pronounced structures, i.e., higher probability of pronounced void regions, to cluster-pronounced structures, yielding higher probability of pronounced cluster regions, along with $St_r$ \cite{K3}. 
Such transition of the spatial structure could be owing to the modulation of clustering production imposed by the conservation of the particle number density $n$.

\end{document}